# Mg(B,O)$_2$ precipitation in MgB$_2$

X. Z. Liao [a)], A. Serquis, Y. T. Zhu, J. Y. Huang, L. Civale, D. E. Peterson,

and F. M. Mueller

Superconductivity Technology Center, Los Alamos National Laboratory

Los Alamos, New Mexico 87545

H. F. Xu

Department of Earth & Planetary Sciences, University of New Mexico

Albuquerque, New Mexico 87131

MgB$_2$ samples prepared by solid-state reaction were investigated using high-resolution transmission electron microscopy (HREM), X-ray energy-dispersive spectroscopy (EDX), electron energy-loss spectroscopy (EELS), and energy-filtered imaging. Large amounts of coherent precipitates with a size range from about 5 nm up to about 100 nm were found in the MgB$_2$ crystallite matrices. The precipitates are of different shapes including sphere, ellipsoid, and faceted polyhedron depending on the size of the precipitates. EDX and EELS analyses confirm that smaller precipitates contain magnesium, boron and oxygen while larger faceted precipitates contain mainly magnesium and oxygen, implying that the oxygen content increases with precipitate size. HREM and electron diffraction investigations found that the precipitates have the same crystal lattice structure as that of MgB$_2$ but with various composition modulations depending on the composition of the precipitates. The precipitates transform to the MgO phase after long exposure to residual oxygen in flowing Ar gas at high temperatures. The effect of the precipitates in different size ranges on flux pinning is discussed.

**PACS numbers**: 74.70.Ad, 74.60.Ge,

---

[a] Electronic mail: xzliao@lanl.gov

I. **Introduction**

The discovery of superconductivity at 39 K in magnesium diboride ($MgB_2$)[1] has been catching the attention of scientists for its possible applications in electro-magnets and in electronic devices. Because of its low cost, low anisotropy, larger coherence lengths, and strong coupling across grain boundaries,[2,3] it may be possible for $MgB_2$ to replace low critical temperature ($T_c$) superconductors in some applications. To make practical devices using $MgB_2$, it is essential to have high critical current density $J_c(H)$ in high magnetic fields H. Significant efforts have been made to improve $J_c$ using various methods, such as irradiation,[4] chemical doping,[5] and increasing grain boundaries through reducing grain size.[6]

It has been reported that chemical doping with oxygen results in very high upper critical fields[7] and increases $J_c$.[8] However, oxygen incorporation into $MgB_2$ also lowers $T_c$.[7,8] In a previous paper,[9] we have presented a way to overcome this negative effect through the segregation of oxygen to form coherent nanometer-sized Mg(B,O) precipitates, which act as effective flux pinning centers. In this way, the $J_c$ of the $MgB_2$ is improved without undermining $T_c$. In this paper, we report a detailed transmission electron microscopy (TEM) investigation on the structures and compositions of the precipitates. We also discuss the effect of the precipitates in different size ranges on flux pinning.

II. **Experimental procedures**

$MgB_2$ samples were prepared by solid-state reaction. As starting materials, amorphous boron powder (-325 mesh, 99.99% Alfa Aesar) and magnesium turnings (99.98% Puratronic) were used in an atomic ratio of magnesium : boron = 1:1. The boron powder was pressed into small pellets with a dimension of 5 mm in diameter and 4 mm in thickness under a pressure of ~500 MPa. All materials were wrapped in Ta foil and then placed in an alumina crucible inside a tube furnace under ultra-high purity (UHP) Ar, heated for two hours at 900°C, slowly cooled down at a rate of 0.5°C/min to 500°C and then furnace cooled to room temperature. More details on the sample preparation have been given elsewhere.[10]

Samples for TEM investigation were prepared by mechanically grinding the $MgB_2$ pellets to a thickness of about 50 µm and then further thinning to a thickness of electron transparency using a Gatan precision ion polishing system with $Ar^+$ accelerating voltage of 3.5 kV. TEM bright-field imaging and electron diffraction investigations were carried out using a Philips CM30 TEM working at 300 kV. High-resolution transmission electron microscopy (HREM) investigations were carried out using a JEOL 3000F TEM working at 300 kV. Composition analysis and elemental mapping were performed in a JEOL 2010F TEM with a field emission gun working at 200 kV. The JEOL 2010F is equipped with an Oxford X-ray energy dispersive spectroscopy (EDX) system and a Gatan image filtering system. The elemental maps were obtained using energy-filtered imaging (EFI) and the three-window technique was used for background removal.[11] The boron K-edge at 188 eV in the electron energy loss spectrum was used for boron mapping, and the centers of two pre-edge windows were set at 148 and 168 eV with a slit width of 20 eV. Oxygen maps were obtained using the oxygen K-edge at 532 eV with the centers of two pre-edge windows set at 492 and 512 eV with a slit width of 20 eV.

To avoid the channeling effect when collecting EDX and electron energy loss spectra (EELS),[12] strong diffraction conditions were avoided by orienting the specimen away from any main zone-axis. Strong diffraction conditions were also prevented in the elemental mapping because the background removal procedure in the three-window technique is unable to totally eliminate the intensity changes caused by diffraction contrast variations that occur between images acquired at different energy losses, leaving artifacts in the final EFI.[13,14]

**III. Experimental results and discussions**

We use the three-index notation system to index the hexagonal $MgB_2$ crystal in this paper. Figure 1 shows a typical [010] zone-axis bright-field TEM image of a $MgB_2$ crystallite. The (001) plane of the crystallite parallels to the vertical edges of the figure and a (100) plane parallels to the horizontal edges of the figure. A large number of precipitates with different shapes and with sizes ranging from about 10 nm to over 100 nm are seen within the $MgB_2$ crystallite. A rough estimation from typical images like Fig. 1 suggests that the densities of the precipitates with size ranges from 10 nm to 50 nm and from 50 nm to 100 nm are ~ $10^{15}/cm^3$ and ~ $10^{14}/cm^3$, respectively. Precipitates with the smallest sizes are spherical (see a typical example marked with "A") while precipitates with medium sizes are elliptical (see a typical example marked with "B"). Precipitates larger than 50 nm show clear facets (see a typical example marked with "D"). Note that the size-range of spherical and elliptical precipitates is partially overlapped. Evidences show that elliptical precipitates could be formed through the coalescence of two closely located spherical precipitates when they grew larger (see a typical example marked with "C").

The factors that influence the shape of a precipitate are interfacial energy and strain energy. As will be discussed later, smaller precipitates have the same lattice structure, a similar lattice parameter (electron diffraction cannot tell the difference in lattice parameter), and the same crystal orientation as the $MgB_2$ matrix. Therefore, strain energy is negligible. Since the crystal lattices of smaller precipitates and the matrix match very well in all interfacial planes, the interfacial energy should not vary much with the orientation of the interface between the smaller precipitates and the matrix. Therefore the smaller precipitates take the spherical shape, which has the smallest interfacial area (and therefore the smallest total interfacial energy). [15] However, with increasing precipitate size, the chemical difference between the precipitates and the $MgB_2$ matrix increases (see discussion on the precipitate compositions later). As a result, it is possible that interfaces along some specific planes [for example, (100)] have lower interfacial energy and the larger precipitates tend to become faceted along these planes.

The composition of the precipitates was investigated using EDX taken from very thin areas at the sample edge. Figure 2 shows the EDX results of (a) the $MgB_2$ matrix, (b) a small precipitate of about 30 nm in diameter, and (c) a large faceted precipitate. Trace amount of oxygen is seen in the $MgB_2$ matrix while the oxygen content is higher in the small precipitate and only magnesium and oxygen signals are seen in the large precipitate.

Because of the small precipitate sizes, signals from the surrounding matrix material might contribute to the EDX spectrum in Fig. 2 (b) and therefore the EDX spectrum itself cannot conclude that the precipitates are made up of magnesium, boron, and oxygen. Because precipitates include much more oxygen than the $MgB_2$ matrix, it is expected that the different chemical environments between the precipitates and the matrix will result in different near edge fine structures in the boron K-edge spectra, if the precipitate does contain boron. To confirm that small precipitates do contain boron, EELS analysis was carried out. The boron K-edge spectra obtained from the $MgB_2$ matrix and a small precipitate are shown in Fig. 3 (a) and (b), respectively. The boron K-edge fine structure in Fig. 3 (a) obtained from $MgB_2$ matrix is very similar to that of $MgB_2$ reported by Klie et al.[16] while clear difference is seen in Fig. 3 (b): the shoulder indicated by an arrow in (a) becomes a small peak in (b), also indicated by an arrow. The difference between the boron K-edge fine structures in Fig. 3 (a) and (b) directly proves that boron exists in the precipitate.

The fact that large precipitates contain mainly magnesium and oxygen while small precipitates contain all three elements magnesium, boron, and oxygen implies that the composition in precipitates varies with precipitate size. The oxygen content increases, while the boron content decreases, with increasing precipitate size, indicating the substitution of boron by oxygen. Unfortunately, it is not possible to quantitatively determine the composition-size relationship because influence from the matrix unavoidably affects the composition investigation of very small precipitates. Quantitative analysis of the EDX spectra in Fig. 2 using magnesium oxide (MgO) located at the $MgB_2$ grain boundaries as the standard sample suggests that the oxygen : magnesium ratio in the $MgB_2$ matrix is about 0.05, while the oxygen : magnesium ratio measured from the small precipitate is about 0.46. Given the fact that smaller precipitates have the same basic lattice structure as the $MgB_2$ matrix (see below), it is expected that the composition of the small precipitate remains the form of $Mg(B,O)_2$. Therefore, the composition of the precipitate that gives the spectrum in Fig. 2 (b) should be $MgB_{1.54}O_{0.46}$. Given the possibility that signals from the surrounding $MgB_2$ matrix material might contribute to the EDX spectrum in Fig. 2 (b), the real oxygen content in the precipitate might be higher than 0.46.

Oxygen in the sample is believed to have come from flowing UHP Ar during the $MgB_2$ synthesis process. A trace amount of oxygen in the UHP Ar, which was used as the protective gas, could dissolve into the $MgB_2$ at high synthesis temperatures. It is believed that $MgB_2$ has relatively high oxygen solubility at the synthesis temperature. When temperature decreased during the cooling process, the solubility of oxygen decreased, forcing the oxygen out of the $MgB_2$ lattice to form $Mg(B,O)_2$ precipitates.

Figure 4 (a) shows a typical elemental map of boron. The dark areas in Fig. 4 (a), which imply less boron content than the matrix, indicate the locations and the sizes of precipitates. Three different precipitate size ranges are seen in Fig. 4 (a): (1) the largest precipitate with a dimension of around 50 nm located at the upper-right corner of the image; (2) three medium-sized precipitates with dimensions of around 10 to 20 nm indicated by three large arrows; and (3) a few very small precipitates with dimensions of about 5 nm forming a precipitate cluster at the upper-left corner of the image, as indicated by small arrows. Note that the small precipitates show very poor image contrast in Fig. 4 (a) and are not seen by the conventional diffraction contrast image in Fig. 1. The reasons contributing to the poor contrast might include: (1) the difference in boron contents between the small precipitates and the surrounding matrix is small and (2) contribution from matrix further smears out the contrast. Figure 4 (b) is a typical elemental map of oxygen taken from the same area as Fig. 4 (a). The areas appearing dark in Fig. 4 (a) are bright in the oxygen map, consistent with the abovementioned conclusion that boron is substituted by oxygen. However, because of the poor intensity, no small precipitate is seen in Fig. 4 (b). The observation in Fig. 4 pushes down the lower limit of observed precipitate dimension from about 10 nm seen in Fig. 1 to about 5 nm. Although there are less very small precipitates than the precipitates with the sizes in the range of 10 ~ 50 nm, the local density of the very small precipitates in some areas, as shown in Fig. 4 (a), is much higher than that of larger precipitates. This implies that ripening (some precipitates grow larger at the expense of others) happened during the growth of these precipitates, much like the precipitation processes observed in other systems.

Large angle tilting electron diffraction experiments on the precipitates suggest that medium- to small-sized precipitates have the same basic lattice structure and orientation as their surrounding $MgB_2$ matrix. No difference in lattice parameters between the precipitates and the matrix can be detected by conventional electron diffraction patterns. However, extra satellite diffraction spots are seen in some directions (see, for example, Fig. 8), implying the structural modulation nature of the precipitates. Most precipitates are more visible and appear darker than the $MgB_2$ matrix in bright-field diffraction contrast images when the $MgB_2$ matrix is under some main zone-axes (see, for example, the [010] image in Fig. 1) imaging conditions. This is because the precipitates have the same orientation as the $MgB_2$ and the substitution of boron by oxygen makes the precipitates stronger in electron scattering.

HREM images taken from various main zone-axes (e.g. [010], [001], [1-11], and [101]) further confirm that the precipitates have the same basic lattice structure as the $MgB_2$ matrix. Typical examples of HREM images of precipitates are shown in Fig. 5. Figure 5 (a) shows an HREM image of a precipitate taken along a [010]. The precipitate is spherical with a diameter of about 15 nm. The precipitate shows the same rectangular lattice fringes as the surrounding matrix. Figure 5 (b) illustrates the [001] HREM image of another precipitate with a diameter of around 30 nm. To see more clearly the lattice fringes of the precipitate and its surrounding matrix area in Fig. 5 (b), an area marked by a white rectangle is enlarged and demonstrated in Fig. 5 (c), which shows exactly the same hexagonal lattice fringes inside and outside the precipitate.

HREM images in Fig. 5 demonstrate that the precipitates have the same *basic* structure as the $MgB_2$ matrix. However, the partial replacement of boron by oxygen in the precipitates does affect the structure of the precipitates as evidenced by extra satellite spots along some directions in the electron diffraction patterns of the precipitates. This is also reflected in HREM images from several directions. Figure 6 (a) shows a [010] HREM image of a precipitate providing evidence of structural modulation in the precipitate. A periodic contrast change is superimposed on two-dimensional lattice fringes, which are the same as the fringes observed in its surrounding $MgB_2$ matrix.[9] The periodic contrast change, appeared as dark-and-bright stripes, is along [100] and the period is five times that of the $MgB_2$ crystal-lattice constant along [100]. This period is further confirmed by Fig. 6 (b), the Fourier transformation of Fig. 6 (a), which shows satellite diffraction spots along $[100]^*$ with a vector equals to $1/5[100]^*$. The fact that the dark and bright stripe areas have the same two-dimensional lattice fringes implies that the crystal structure at these areas remains the same. The periodic contrast difference is caused by periodic composition (oxygen and boron) variation, with the dark areas having higher oxygen content[17]. Therefore, the precipitates are of the basic $MgB_2$ crystal structure with compositional modulation.

Note that Moiré fringes generated by two overlapping crystals can also present similar image contrast as shown in Fig. 6 (a). However, 20% lattice mismatch between the two overlapping crystals is needed to have Moiré fringes with the period shown in Fig. 6 (a). The 20% lattice mismatch is much larger than the very small lattice difference between the precipitate and the matrix. Furthermore, a typical HREM image of Moiré fringes normally appears differently at the bright and dark areas (see, for example, ref. 18), which is not the case in Fig. 6 (a).

Figure 7 (a) is another example of a [010] HREM image showing compositional modulation superimposing on the basic structure of $MgB_2$. A spherical precipitate with a diameter close to 15 nm demonstrates slightly darker contrast than its surrounding matrix in Fig. 7 (a). The precipitate/matrix boundary is highlighted using white stars and an area marked with a white rectangle is enlarged and inserted on the upper right corner of Fig. 7 (a), showing clearly the same basic structure between the precipitate and the matrix. However, dim stripes are seen within the precipitate along the direction indicated by a white line drawn within the precipitate. Fourier transformations of the image from the matrix area and from the precipitate are shown in Fig. 7 (b) and (c), respectively. Besides the basic spots shown in Fig. 7 (b), extra satellite spots are seen in Fig. 7 (c) with a vector of about [-0.26, 0, 0.39]*.

Obviously, the modulations shown in Fig. 6 and Fig. 7 are different. In fact, more than one type of modulated structures have been found in precipitates. An evidence is shown in a [010] electron diffraction pattern in Fig. 8, where extra satellite diffraction sports along the [001]* with a vector of about 2/5[001]*are seen. Because the [001]* is unique in the hexagonal $MgB_2$ structure and because no extra satellite spot along [001]* is seen in Figs. 6 and 7, it is concluded that the diffraction pattern shown in Fig. 8 comes from a structure that is different from the structure(s) that give rise to the Figs. 6 and 7. It is believed that different oxygen contents are responsible for the different modulations in precipitates.

It is found that the synthesis parameters significantly affect the density and structure of the precipitates. Longer oxygen exposure time at the high temperature of 900ºC increases the density of the precipitates but no significant change in the average precipitate size has been observed. This is beneficial for improving flux pinning. However, longer oxygen exposure time also increases the amount of MgO at grain boundaries, which acts as "weak link" and therefore is undesirable. Therefore, it is important to balance between good flux pinning and good grain boundary connections in choosing synthesis parameters. Longer exposure to trace amount of oxygen at high temperatures also results in the transformation of the precipitates from the hexagonal $Mg(B,O)_2$ phase with compositional modulations to the face-centered cubic MgO phase, implying $Mg(B,O)_2$ is not a stable phase at high temperature in an environment containing trace amount of oxygen. The orientation relationship between the resulting MgO precipitates and the $MgB_2$ matrix revealed by an HREM image and its Fourier transformation in Fig. 9 is: $[001]_{MgB2}$ // $[111]_{MgO}$, and $(110)_{MgB2}//(1\text{-}10)_{MgO}$ (note that planar distance $d(110)_{MgB2} = 0.154$ nm while $d(220)_{MgO} = 0.149$ nm. The lattice mismatch between $d(110)_{MgB2}$ and $d(220)_{MgO}$ is only 3.2%). In fact, Zhu et al.[19] have also reported that the precipitates in their $MgB_2$ samples are MgO.

A possible reason for the small MgO precipitate sizes after longer time high temperature annealing (ripening did not happen) is that MgO is a very stable compound.

Once it forms, the energy barrier for it to dissociate for the oxygen to migrate to form large precipitates is very high.

**IV. Effect on vortex pinning**

Precipitates in a superconductor can pin vortices[20,21,22] by reduction or suppression of the condensation energy within their volume, and more indirectly by creating defects such as stalking faults, dislocations and microcracks in the superconducting matrix. This second effect is relevant, for instance, in the case of inclusions of $Y_2BaCuO_5$ (the "211" phase) in $YBa_2Cu_3O_7$. In our samples, in contrast, the precipitates are coherent with the matrix lattice, with a remarkable absence of surrounding matrix defects, so pinning is likely to arise only from the first mechanism. Thus, the elementary pinning energy of one defect, $u_p$, is given by the difference between the condensation energy of a vortex core sitting on top of the defect and far away from it, $u_p = \eta(H_c^2/8\pi)V$, where $\eta \leq 1$ indicates the fraction of suppression of the order parameter ($\eta = 1$ for a non-superconducting precipitate), V is the pinned volume of the core, and $H_c = \Phi_0/(2\sqrt{2}\pi\xi\lambda)$ is the thermodynamic critical field. Here $\Phi_0$ is the flux quantum, $\xi$ is the coherence length, and $\lambda$ is the penetration depth. For simplicity, we will assume that the defects are spherical, which is almost exactly the case in our sample. If the radius of the precipitate, R, is smaller than the radius of the vortex core, $\sqrt{2}\xi$, then $V = (4\pi/3)R^3$, else $V = 2\pi\xi^2 R$. The elementary pinning force $f_p$ is the maximum gradient in $u_p$, and it can be estimated as $u_p$ divided by the distance over which the order parameter recovers. Taking into account the sharpness of the defect-matrix interfaces, we can assume that even for $R > \sqrt{2}\xi$ such recovery takes place over the shortest possible distance, namely $\sim \xi$. (This is equivalent to say that in the large defects pinning occurs essentially at the interface.) We thus obtain $f_p = (\eta/6)H_c^2 R^3/\xi$ for $R < \sqrt{2}\xi$ and $f_p = (\eta/4)H_c^2 R\xi$ for $R > \sqrt{2}\xi$.

The critical current density is $J_c = cF_p/B$, where c is the speed of light, $F_p$ is the pinning force per unit volume, and B is the magnetic induction. In general, the relation between $F_p$ and $f_p$ is highly nontrivial. However, for very low defect density we can make a crude estimate of $F_p$ by simply summing up the elementary pinning forces. This is clearly an overestimate, as both the cost in elastic energy and the partial cancellation of the forces produced by the random distribution of defects will tend to reduce $F_p$. In this approximation, the contribution of defects of a certain radius is $F_p(R) \sim f_p(R)N_p(R)$, where $N_p(R)$ is the number of vortex-defect interactions per unit volume. Let us first consider that R is much smaller than the inter-vortex distance $a = 1.075(\Phi_0/B)^{1/2}$. In this limit, $N_p$ can be approximated as the fraction $2\pi\xi^2 B/\Phi_0$ of the total number of defects of that size per unit volume, $n_p(R)$, that lay within the cores. By combining the above relations we can write the critical current density due to defects of radius R as $J_c(R) \sim (\sqrt{3}/2)\pi\eta(R)[RSn_p(R)]J_0$, where $S = R^2$ for $R < \sqrt{2}\xi$ and $S = (3/2)\xi^2$ for $R > \sqrt{2}\xi$, and where to facilitate the numerical comparisons we have introduced the depairing current density $J_0 = cH_c/(3\sqrt{6}\pi\lambda)$. For T=0, using[23] $\lambda \sim 110$ nm and $\xi \sim 4$ nm, we obtain $H_c \sim 5100$ Oe and $J_0 \sim 2\times10^8$ A/cm$^2$.

As mentioned before, the density of the precipitates with the size range of 10 nm - 50 nm, or 5 nm < R < 25 nm, is $\sim 10^{15}$ defects/cm$^3$. We also know that in defects of these sizes a large fraction of the boron has been replaced by oxygen, so we can assume that the superconductivity is very strongly suppressed in them, thus $\eta \sim 1$. Taking an average radius R = 15 nm we get $J_c(15$ nm$) \sim 2\times10^5$ A/cm$^2$. In a previous study we had reported that, at 5

K and H ~ 0, this particular sample has $J_c$ ~ $5 \times 10^5$ A/cm$^2$.[24] Considering the crudeness of our estimates, we conclude that these defects are good candidates to be one of the main sources of vortex pinning in our samples, and the controlled increase in their density may result in an increase in $J_c$.

In contrast, the smallest defects are probably rather inefficient. It is of course a general rule that defects much smaller than $\xi$ are poor vortex pins. Taking R ~ 2.5 nm, we find $J_c$(2.5nm) ~ $8.5 \times 10^{-12}$ $\eta$(2.5nm) $n_p$(2.5nm) A/cm$^2$. The density of these defects is hard to estimate, as they are difficult to see in the TEM images. We can take the inverse approach and find that, even assuming $\eta = 1$, we need $n_p$ ~ $2 \times 10^{16}$ cm$^{-3}$ to obtain a contribution similar to the larger defects. This $n_p$ is too large, certainly well above the observed density. Moreover, it corresponds to an average distance between defects of ~ 30nm, so small as to invalidate the very low defect density approximation. On top of that, these small defects have a much smaller oxygen content, suggesting that $\eta \ll 1$ and thus further reducing $J_c$.

Finally, we have to consider the largest defects. We observe about $10^{14}$ defects/cm$^3$ with 25 nm < R < 50 nm. Taking R = 40 nm we obtain $J_c$(40 nm) ~ $5 \times 10^4$ A/cm$^2$, smaller than the contribution of the "mid size" defects, but still significant. On the other hand, for fields above ~ 3 KOe the vortex spacing becomes smaller than the diameter of these defects, so each one can pin more than one vortex. Thus, these large defects may become comparatively more important as H increases. Detailed studies of the field and temperature dependence of pinning are required to clarify this point. Additional vortex pinning may arise from the grain boundaries,[2] but the analysis of that contribution is beyond the scope of the present study.

## V. Summary

A large amount of nanometer-sized coherent precipitates were found in $MgB_2$ matrix prepared by solid-state reaction synthesis. Compositional analysis suggests that the precipitates are made up of magnesium, boron, and oxygen. The oxygen content increases while boron content decreases with increasing precipitate sizes. It is believed that oxygen was introduced into the $MgB_2$ sample during high temperature synthesis. The oxygen, dissolved into $MgB_2$ at high temperature, was later forced out to form $Mg(B,O)_2$ precipitates due to its lower solubility at lower temperatures. The precipitates are of the same basic structure as the $MgB_2$ matrix but with composition modulations. Long time exposure to oxygen at high temperatures results in the transformation of $Mg(B,O)_2$ precipitates to MgO with little change in precipitate sizes. Those precipitates with radius larger than ~ 5 nm (or diameters larger than ~ 10 nm) are relevant pinning centers in $MgB_2$.

## Acknowledgments

This work was performed under the auspices of the US Department of Energy, Office of Energy Efficiency and Renewable Energy, as part of its Superconductivity for Electric Systems program. The JEOL 2010F in the TEM Laboratory in the Department of Earth and Planetary Sciences of the University of New Mexico is funded by NSF.

**Figure captions**

Figure 1 [010] zone-axis bright-field diffraction contrast images of a $MgB_2$ crystallite. High density of precipitates with different shapes and sizes are clearly seen. Some precipitates are labeled with A, B, C, and D and indicated with arrows, demonstrating the change of shape with size and coalescence during the growth of the precipitates.

Figure 2 EDX spectra of (a) the $MgB_2$ matrix, (b) a small precipitate, and (c) a large faceted precipitate.

Figure 3 EELS spectra taken from (a) the $MgB_2$ matrix and (b) a small precipitate showing difference in the detailed shape of boron K-edge. The position indicated by an arrow has changed from a shoulder in (a) to a peak in (b).

Figure 4 EFI images of (a) a boron map and (b) an oxygen map. Large arrows indicate precipitates with size range of 10 to 20 nm. Small arrows indicate precipitates in the size range of around 5 nm.

Figure 5 HREM images of precipitates taken from (a) a [010] direction and (b) the [001] direction. The area marked with a white rectangle in (b) is enlarged and shown in (c).

Figure 6 (a) A [010] HREM image showing basic lattice fringes, which are the same as the $MgB_2$ matrix, and periodic contrast change marked with black bars; (b) Fourier transformation of (a). Satellite diffraction spots are seen along [100]$^*$ with a period of 1/5[100]$^*$.

Figure 7 (a) An HREM image along [010] direction. A spherical precipitate with a diameter of about 15 nm and with darker contrast than the MgB$_2$ matrix is clearly seen. The precipitate boundary is highlighted using white stars. Weak stripe contrast along the direction marked by a white line is seen in the precipitate. A magnified image at the precipitate-matrix boundary marked by a white rectangle is shown at the upper right corner, which attests to the coherent nature of the precipitate and the same basic structure of the matrix and the precipitate. (b) Fourier transformation of the matrix. (c) Fourier transformation of the precipitate showing satellite spots with a vector of [-0.26, 0, 0.39]*.

Figure 8 A [010] diffraction pattern of a precipitate showing modulation satellite spots along [001]* with a period of roughly 2/5[001]*.

Figure 9 (a) An HREM image of a MgO precipitate and the surrounding MgB$_2$ matrix (shown in a lower magnification) and (b) its Fourier transformation showing the orientation relationship between the MgO precipitate and it MgB$_2$ matrix.

Figure 1

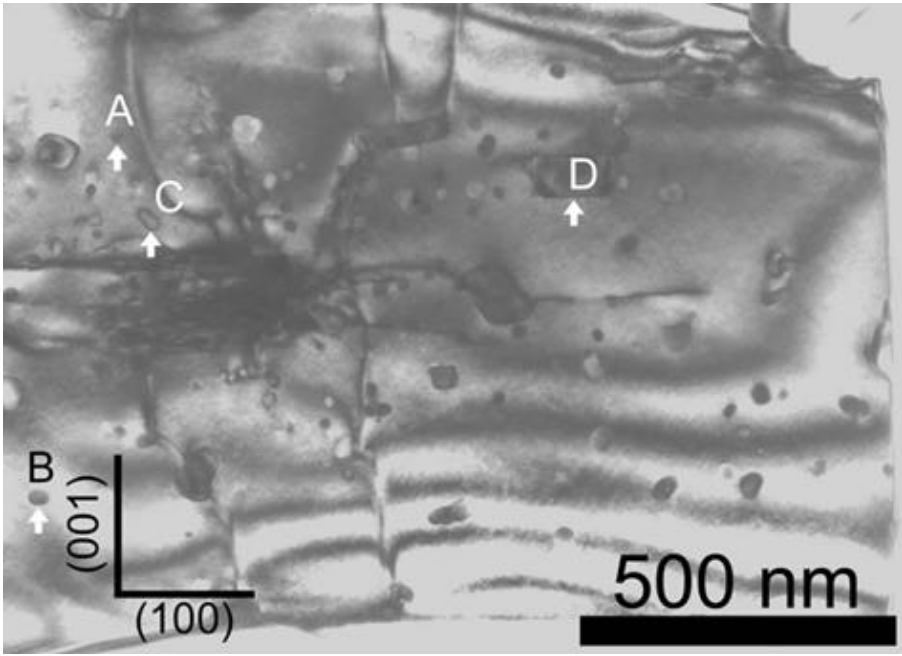

Figure 2

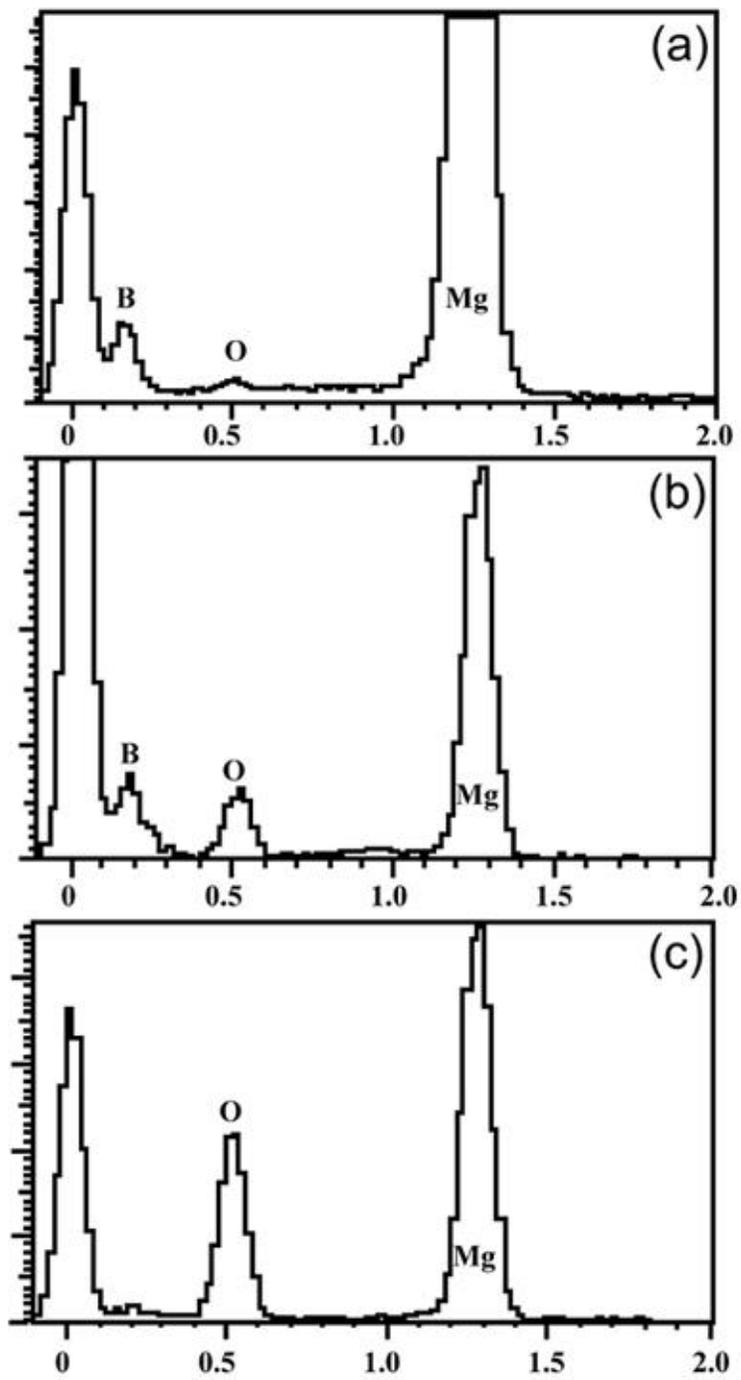

Figure 3

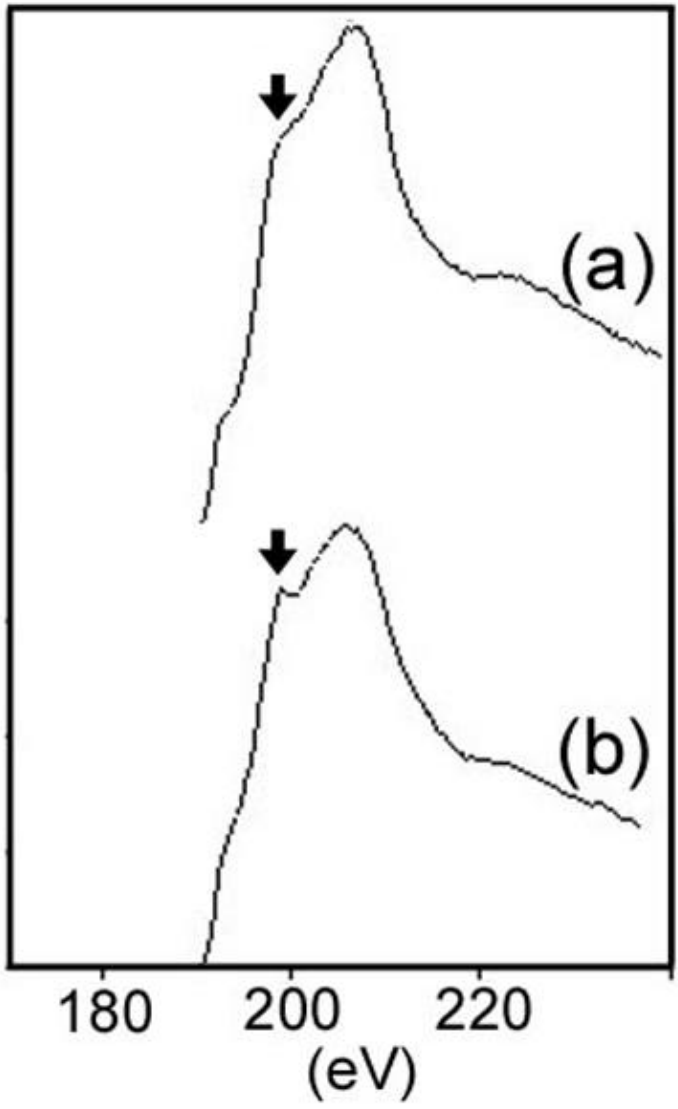

Figure 4

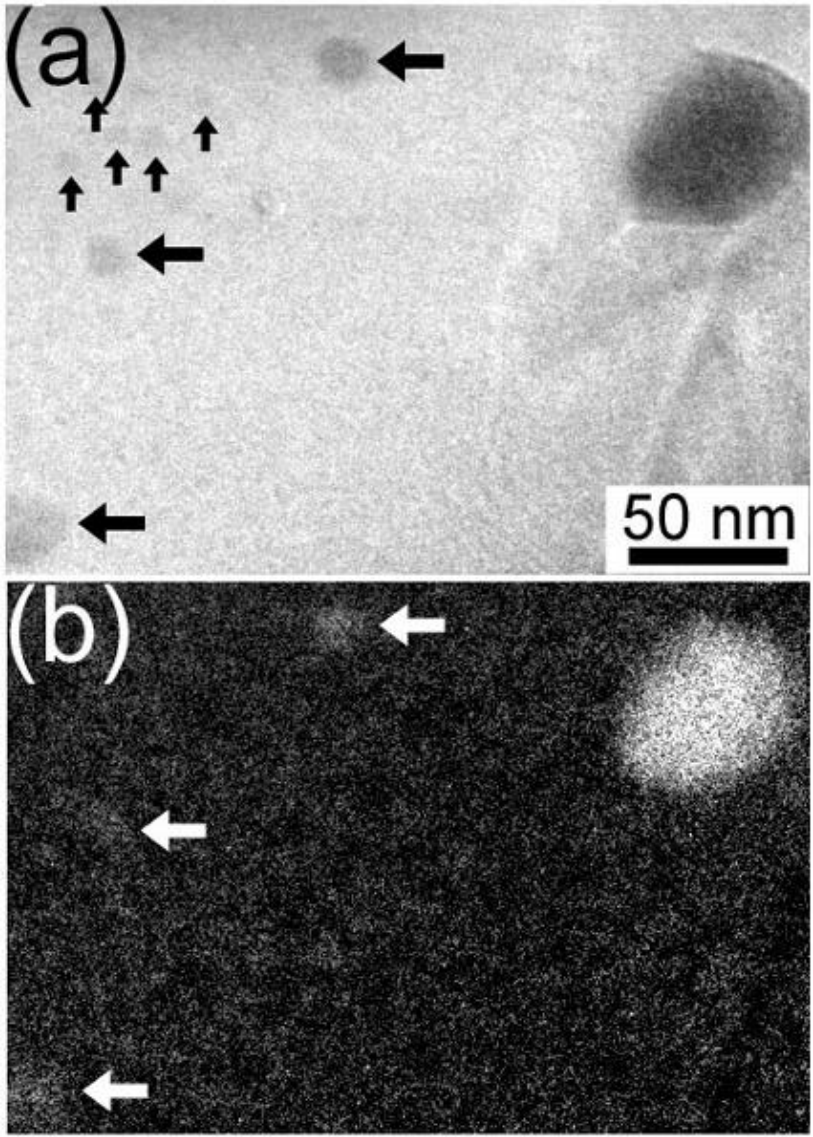

Figure 5

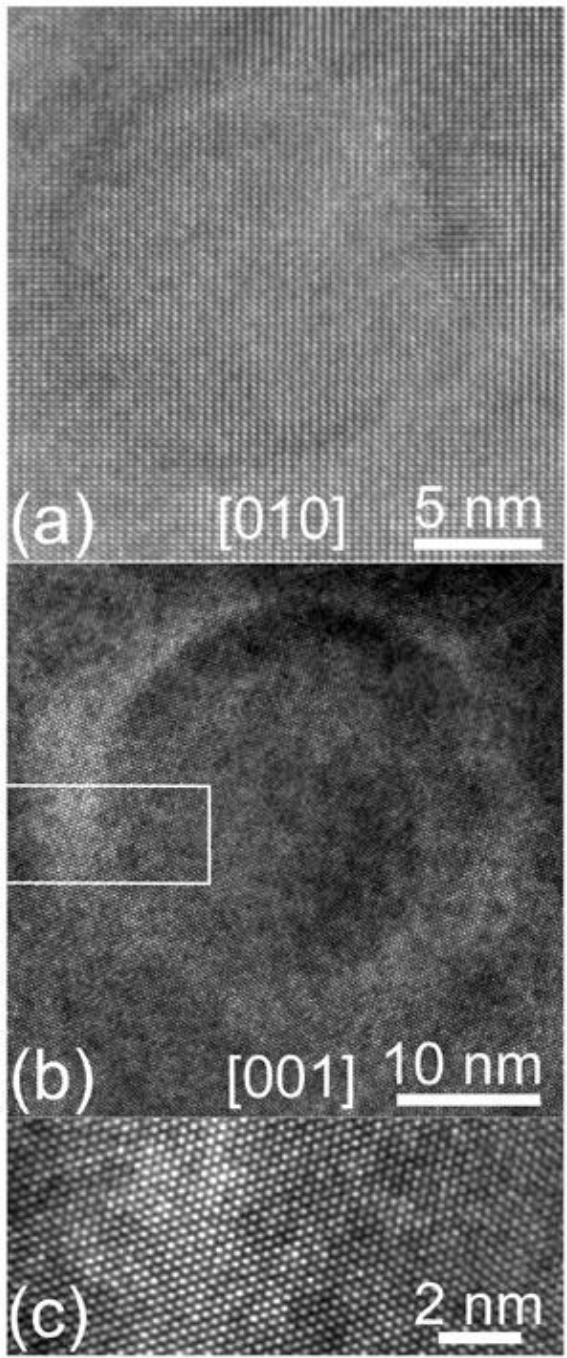

Figure 6

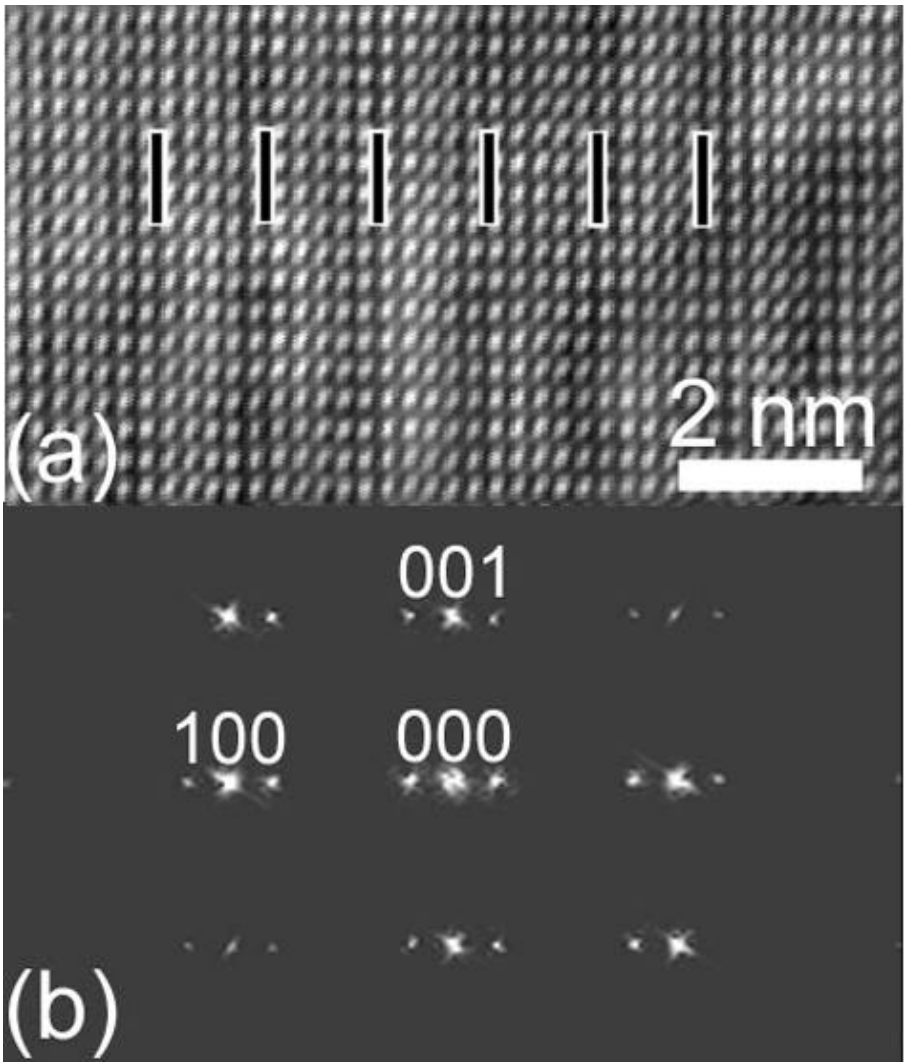

Figure 7

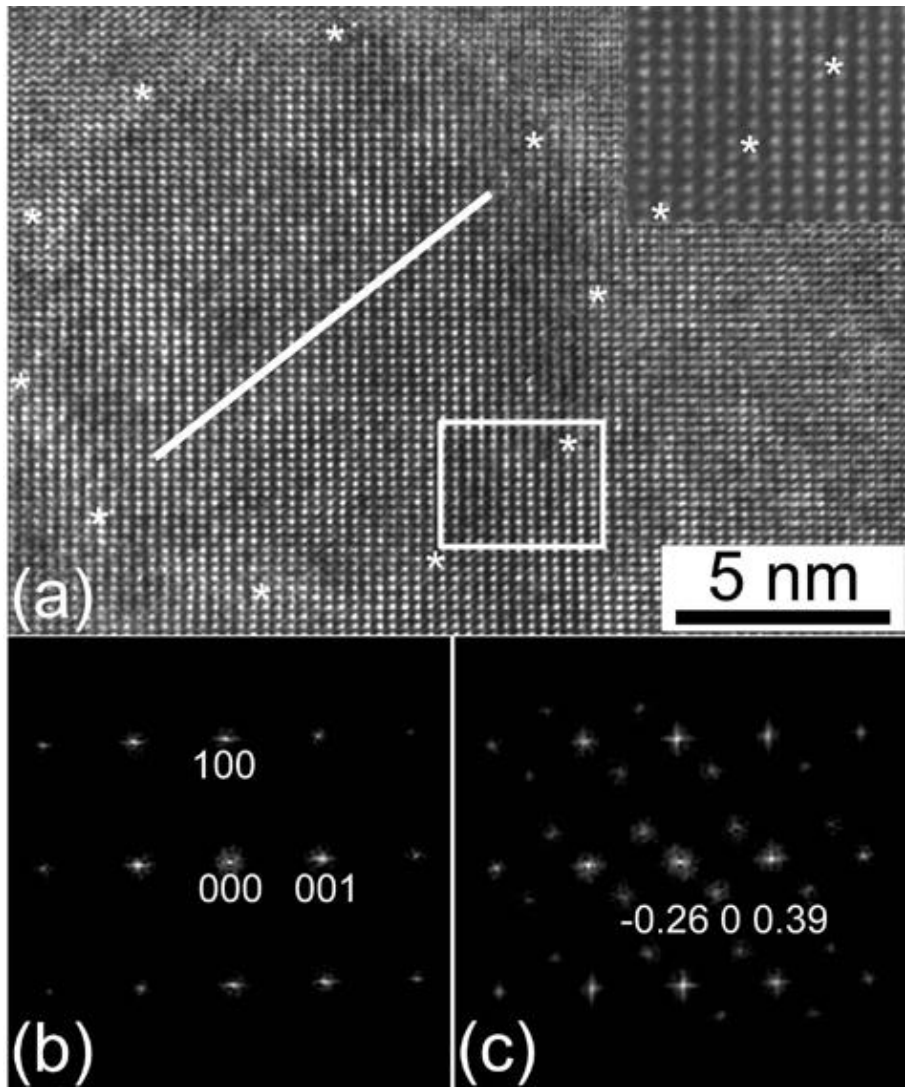

Figure 8

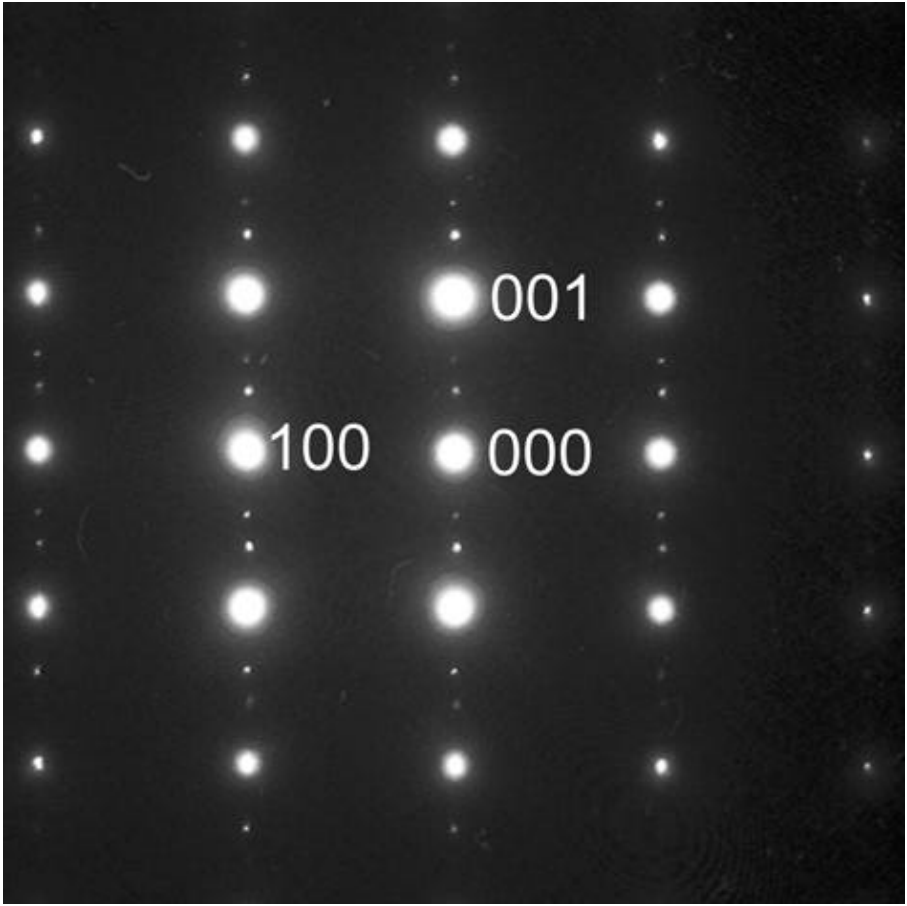

Figure 9

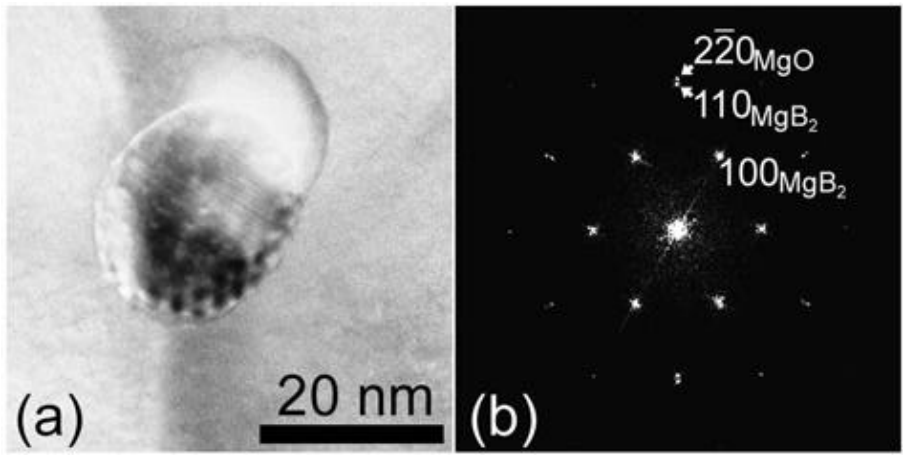